\documentclass[a4paper,11pt]{article}
\usepackage{pos}

\title{An end-to-end calibration of the Mini-EUSO detector in space}
\ShortTitle{MiniEUSOflasher}

\author*[a,b]{Hiroko Miyamoto}
\author[b]{Matteo Battisti}
\author[b,c,d]{Dario Barghini}
\author[e,f]{Alexander Belov}
\author[b,c]{Mario Bertaina}
\author[c]{Marta Bianciotto}
\author[g]{Francesca Bisconti}
\author[h]{Carl Blaksley}
\author[i]{Sylvie Blin}
\author[j]{Karl Bolmgren}
\author[g,k]{Giorgio Cambi\`{e}}
\author[l]{Francesca Capel}
\author[g,h,k]{Marco Casolino}
\author[m]{Igor Churilo}
\author[o]{Christophe De La taille}
\author[h]{Toshikazu Ebisuzaki}
\author[p]{Johannes Eser}
\author[q]{Francesco Fenu}
\author[r]{Geroge Filippatos}
\author[s]{Massimo Alberto Franceschi}
\author[j]{Christer Fuglesang}
\author[b,c]{Alessio Golzio}
\author[t]{Philippe Gorodetzky}
\author[t]{Fumioshi Kajino}
\author[h]{Hiroshi Kasuga}
\author[f]{Pavel Klimov}
\author[r]{Viktoria Kungel}
\author[m]{Vladimir Kuznetsov}
\author[b,c]{Massimiliano Manfrin}
\author[g]{Laura Marcelli}
\author[n]{Gabriele Mascetti}
\author[u]{W{\l}odzimierz Marsza{\l}}
\author[b]{Marco Mignone}
\author[f]{Alexey Murashov}
\author[s]{Tommaso Napolitano}
\author[h]{Hitoshi Ohmori}
\author[p]{Angela Olinto}
\author[i]{Etienne Parizot}
\author[g,k]{Piergiorgio Picozza}
\author[v]{Lech Wiktor Piotrowski}
\author[b,c]{Zbigniew Plebaniak}
\author[i]{Guillaume Pr\'{e}v\^{o}t}
\author[g,k]{Enzo Reali}
\author[s]{Marco Ricci}
\author[g,k]{Giulia Romoli}
\author[h]{Naoto Sakaki}
\author[f]{Sergei Sharakin}
\author[u]{Kenji Shinozaki}
\author[u]{Jacek Szabelski}
\author[h]{Yoshiyuki Takizawa}
\author[n]{Giovanni Valentini}
\author[u]{Michal Vrabel}
\author[r]{Lawrence Wiencke}
\author[f]{Mikhail Zotov}

\affiliation[a]{Gran Sasso Science Institute,
  Viale Francesco Crispi 7, 67100 l'Aquila, Italy}

\affiliation[b]{INFN Section of Turin,
  Via P. Giuria 1, 10125 Turin, Italy}

\affiliation[c]{University of Turin, Department of Physics,
  Via P. Giuria 1, 10125 Turin, Italy}

\affiliation[d]{INAF Astrophysics Observatory of Turin,
  Via Ossevatorio 20, 10025 Pino Torinese, Italy}

\affiliation[e]{M.V.Lomonosov Moscow State Univ., Faculty of Physics, 
  Leninskie gory 1(2), 119234 Moscow, Russia}

\affiliation[f]{M.V.Lomonosov Moscow State Univ., Skobeltsyn Institute of Nuclear Physics, 
  Leninskie gory 1(2), 119234 Moscow, Russia}

\affiliation[g]{INFN Section of Rome Tor Vergata,
  Via della Ricerca Scientifica 1, 00133 Rome, Italy}

\affiliation[h]{RIKEN,
  2-1 Hirosawa Wako, 351-0198 Saitama, Japan}

\affiliation[i]{Universt\'e Paris Cit\'e, CNRS, Astroparticule et Cosmologie
  10 Rue Alice Domon et \`{e}onie Duquet, 75013 Paris, France}

\affiliation[j]{KTH Royal Institute of Technology,
  Brinellv\"{a}gen 8, 114 28 Stockholm, Sweden}

\affiliation[k]{University of Rome Tor Vergata, Department of Physics,
  Via della Ricerca Scientifica 1, 00133 Rome, Italy}

\affiliation[l]{Technical University of Munich,
  Arcisstra{\ss}e 21, 80333 Munich, Germany}

\affiliation[m]{S.P. Korolev Rocket and Space Corporation Energia,
  Lenin str., 4a Korolev, 141070 Moscow area, Russia}

\affiliation[n]{ASI, Italian Space Agency,
  Via del Politecnico, 00133 Rome, Italy}

\affiliation[o]{Omega, Ecole Polytechique, CNRS/IN2P3,
  Rte de Saclay, 91120 Palaiseau, France}

\affiliation[p]{They University of Chicago, Department of Astronomy and Astrophysics,
  5640 S. Ellis Avenue, 60637 Chicago IL, US}

\affiliation[q]{Karlsruhe Institute of Technology,
  Hermann-von-Helmholtz-Plats 1, 76344 Eggenstein-Leopoldshafen, Germany}

\affiliation[r]{Colorado Schoold of Mines, Department of Physics,
  1523 Illinois St., 80401 Golden CO, US}

\affiliation[s]{INFN National Laboratories of Frascati,
  Via Enrico Fermi 54, 00044 Frascati, Italy}

\affiliation[t]{Konan University, Department of Physics,
  8 chome-9-1 Okamoto, Higashinada Ward Kobe, 658-8501 Hyogo, Japan}

\affiliation[u]{National Centre for Nuclear Researche,
  28 pu{\l}ku Strzelc\'{o}w, Kaniowskich 69, 90-558 {\L}\'{o}d\'{z}, Poland}

\affiliation[v]{University of Warsaw, Faculty of Physics,
  Ludwika Pasteura 5, 02-093 Warsaw, Poland}

\onbehalf{for the JEM-EUSO collaboration} 

\emailAdd{miyamoto@to.infn.it}
\emailAdd{battisti@to.infn.it}

\abstract{
Mini-EUSO is a wide Field-of-View (FoV, 44$^{\circ}$) telescope currently in operation from a nadia-facing UV-transparent window in the Russian Zvezda module on the International Space Station (ISS). It is the first detector of the JEM-EUSO program deployed on the ISS, launched in August 2019. The main goal of Mini-EUSO is to measure the UV emissions from the ground and
atmosphere, using an orbital platform. Mini-EUSO is mainly sensitive in the 290-430 nm bandwidth. Light is focused by a system of two Fresnel lenses of 25 cm diameter each on the Photo- Detector-Module (PDM), which consists of an array of 36 Multi-Anode Photomultiplier Tubes (MAPMTs), for a total of 2304 pixels working in photon counting mode, in three different time
resolutions of 2.5 ${\mu}$s, 320 ${\mu}$s, 40.96 ms operation in parallel. In the longest time scale, the data is continuously acquired to monitor the UV emission of the Earth. It is best suited for the
observation of ground sources and therefore has been used for the observational campaigns of the Mini-EUSO. In this contribution, we present the assembled UV flasher, the operation of the field campaign and the analysis of the obtained data. The result is compared with the overall efficiency computed from the expectations which takes into account the atmospheric attenuation and the parameterization of different effects such as the optics efficiency, the MAPMT detection efficiency, BG3 filter transmittance and the transparency of the ISS window.
}

\ConferenceLogo{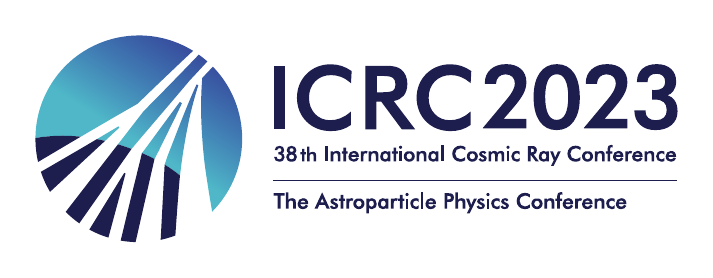}

\FullConference{%
38th International Cosmic Ray Conference (ICRC2023)\\
  26 July - 3 August, 2023\\
  Nagoya, Japan}

\begin{document}
\maketitle

\section{Introduction}
\label{intro}
\begin{figure}[h]
  \centering
  \includegraphics[width=8.4cm,height=4.6cm]{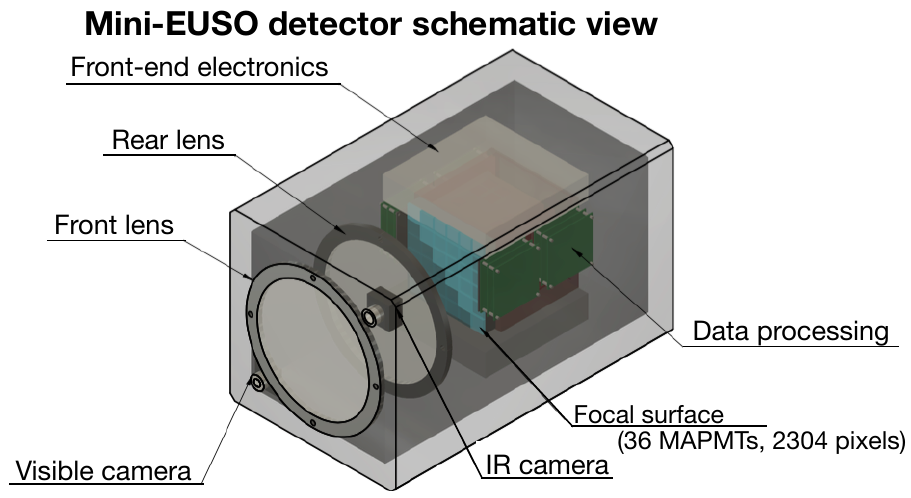}
  \includegraphics[width=6.6cm,height=4.6cm]{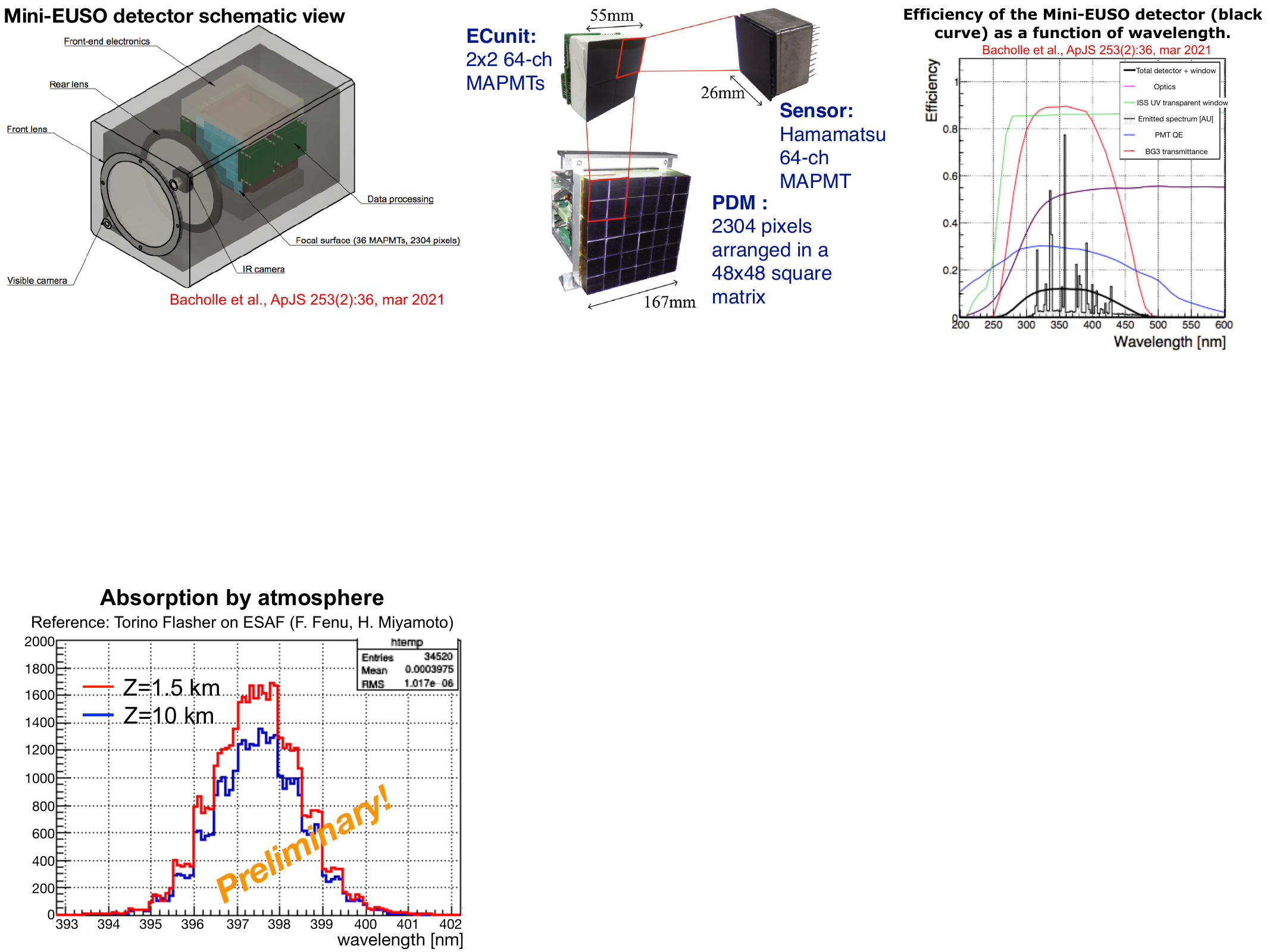}        
  \vspace{-0.2cm}  
  \caption{
    Schematic view of Mini-EUSO telescope (left) and the PDM (right).
    \vspace{-0.4cm}
  }
  \label{fig:ME}       
\end{figure}

Mini-EUSO, a fluorescence telescope with a wide FoV of 44$^{\circ}$ currently in operation
onboard the ISS, is the first detector in space of the JEM-EUSO program~\cite{ref:jemeuso},
launched in August 2019.
The
main goal of Mini-EUSO is to measure the UV emissions from the ground and atmosphere and to assess
the feasibility and performance of the measurement of ultra-high energy cosmic rays by means of a 
space-based detector.
Fig.~\ref{fig:ME} shows schematic views of the full Mini-EUSO telescope (left)
and the PDM (right).
A more detailed explanation of the Mini-EUSO
detector and data acquisition chain is reported in~\cite{ref:MElaunch}.

Several kinds of ground-based flashers have been developed in different groups of the JEM-EUSO collaboration such as in Japan, Italy and France.
One of such flashers has been developed in Turin (namely ``Torino UV flasher''), and calibrated with the Torino EC telescope, a telescope based on a Mini-EUSO EC and electronics. 
The flasher consists of an array of 9 100W COB-UV LEDs, batteries and an Arduino circuit (see the left panel of Fig.~\ref{fig:measUVflasher}).
%
LEDs were programmed to pulse 6 times in 12 s with a pulse duration of 1600 ms on and 400 ms off each, followed by 12 pulses in 9.6 s with 400 ms on and 400 ms off each. 
The durations were decided taking into account that a pixel FoV of Mini-EUSO moves completely to neighboring pixel every $\sim$800 ms as it corresponds to $\sim$6.3 km on ground and the ISS speed is $\sim$7.5 km/s.
In this way, it was 
guaranteed to have a light signal lasting the entire transit of the flasher in a pixel FoV 
(1600 ms on)
and the possibility to measure 
the flasher light and the background (400 ms on).

In May 2021, an observational campaign was performed at Piana di Castelluccio in central Italy at the altitude of $\sim1550$~m a.s.l.
The place was chosen based on the very low light pollution in an area of several km radius.
In this campaign, photons from the Torino UV flasher were detected and we preliminarily estimated the overall efficiency of Mini-EUSO telescope. 
However, the flasher signal was detected by Mini-EUSO at the edge of the FoV of an MAPMT.
Therefore, some light was missing as it was focused in the gap between two PMTs, which is making the calibration effort more uncertain.

Prior to the measurement by Mini-EUSO, the flasher was tested with the Torino EC telescope
at the TurLab facility~\cite{ref:TurLab}. 
As shown in the right panel of Fig.~\ref{fig:measUVflasher}, this facility allows one to place the telescope at 40~m distance from a light source in a dark environment.

An EC based telescope, so-called ''Torino EC telescope" (Fig.2), has been originally built in Turin for fundamental functionality tests, the study and development of the trigger for the Mini-EUSO telescope at the TurLab facility and open-sky conditions~\cite{ref:MEEM}.
It consists of a 30 cm lens tube with a 1-inch plano-convex lens, an Elementary Cell unit (ECunit), which consists of the 2$\times$2 MAPMTs, front-end electronics based on SPAICROC3 ASICs (EC\_ASIC), and the Zynq Board connected to a PC via ethernet cable, where dedicated software for the Mini-EUSO data processing system is installed.
MAPMTs and electronics boards are powered by external High, and Low Voltage DC Power Supply (HVPS, LVPS), respectively.

Fig.\ref{fig:lightcurve} shows examples of the Torino flasher data taken by the Torino EC telescope (top) and by the Mini-EUSO detector in space (bottom), where one can see the image of one frame (integration of 40.96 ms) on the left and the light curves of the UV flasher signal (right) after the background subtraction.

The overall efficiency of Mini-EUSO telescope can be estimated as:
\begin{equation}
  \qquad\mathit{Eff} = \mathit{N_\mathrm{det}} / N{\substack{window\\photons}},    
    \label{eq:eff}
\end{equation}
where $\mathit{N_\mathrm{det}}$
is the photon counts detected by Mini-EUSO telescope and
$N\substack{window\\photons}$
is the number of photons arriving at the UV transparent window on the ISS.
For the analysis, we selected two pixels, pixel[31,5] and pixel[31,13], 
which are located at $18^\circ$ and $12^\circ$ from nadir, respectively,
where the flasher is flashing in the
center of a pixel according to the following methods for each:
\begin{itemize}
  \renewcommand{\labelitemi}{}
\item - method 1: Search for a pixel with flasher signal when the
  neighboring
  top, bottom and left pixels have 
the same counts (the right pixel belongs to a different MAPMT and gaps exist between MAPMTs).
  \label{item:method1}  
\item - method 2: Estimate the geographical position
  from the 
  orbital information (UTC time, distance and the angle)
  \label{item:method2}
\end{itemize}
As a result, 
$\mathit{N_\mathrm{det}}$
are 9.7 cts/GTU (pix[31,5]) and 8.3 cts/GTU (pix[31,13]), respectively.
In the following we describe how
$N\substack{window\\photons}$
has been estimated.
\begin{figure}[h]
  \centering
  \vspace{-0.2cm}  
  \includegraphics[width=12cm,clip]{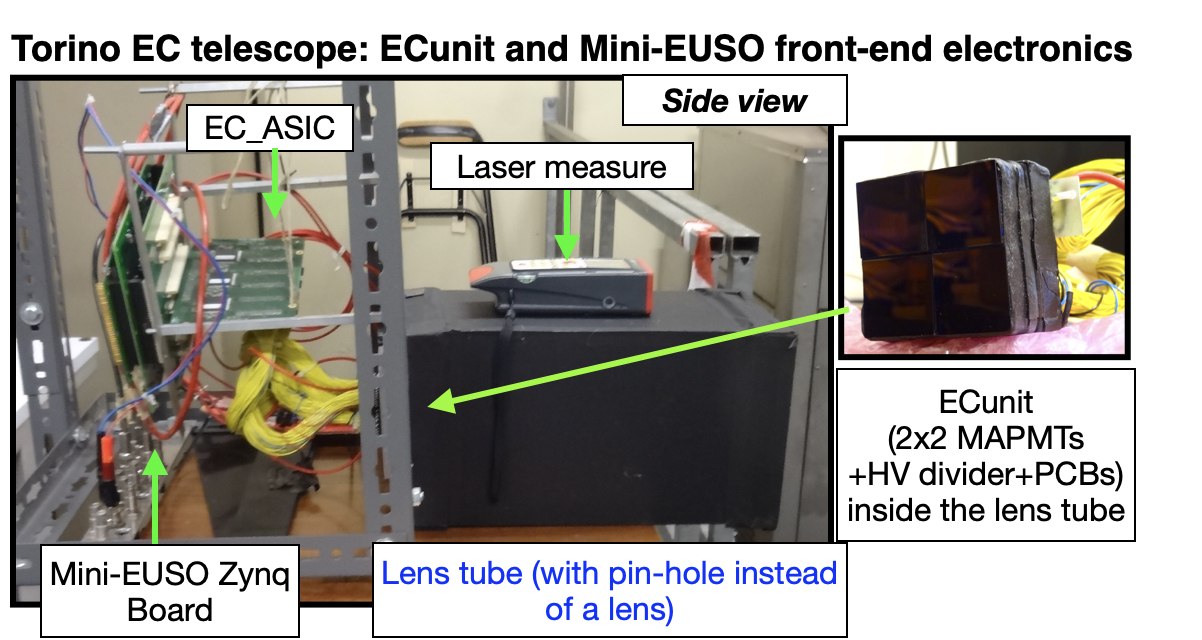}    
  \caption{
    The Torino EC Telescope consists of an ECunit, EC\_ASIC, Zynq Board, lens tube, 1'' plano-convex lens, 
CPU (PC), external LV and HV power supplies.
    For the flasher measurements in the lab, a pin-hole of 0.1~mm in diameter, instead of the lens, is used to reduce light from the flasher LEDs.
    \vspace{-0.8cm}
  }
  \label{fig:TorinoEC}
\end{figure}
%
%
%
%
\begin{figure*}[h]
  \begin{center}
  \includegraphics[width=13cm,clip]{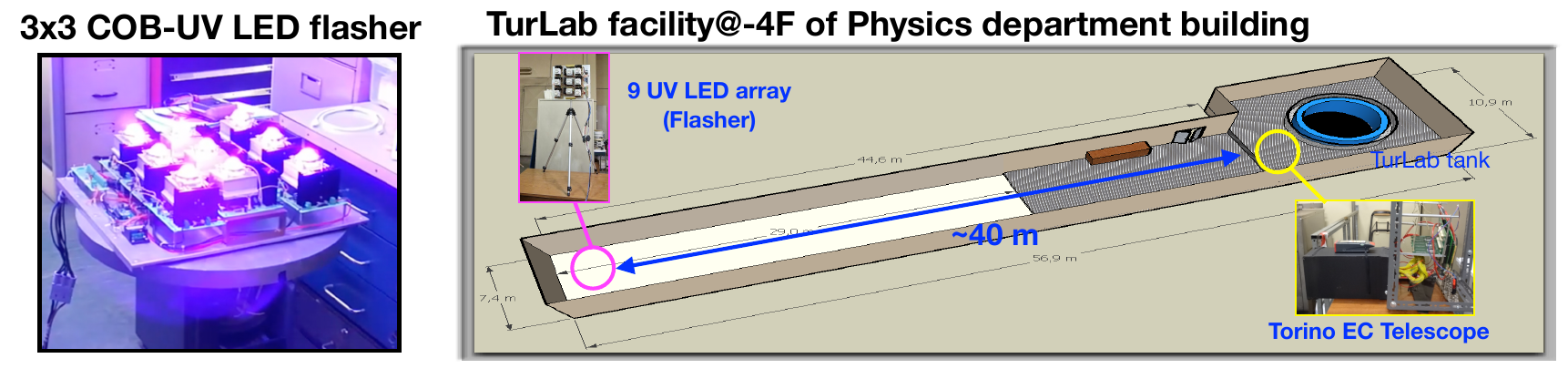}
  \caption{
    Left: $3\times3$ COB-UV LED flasher built in Turin.
    Right: schematic view of TurLab facility where the positions of the flasher and Torino EC telescope are indicated.
  }
  \label{fig:measUVflasher}
  \end{center}
\end{figure*}
\begin{figure*}[h]
  \begin{center}
    \includegraphics[width=12cm,height=6.4cm]{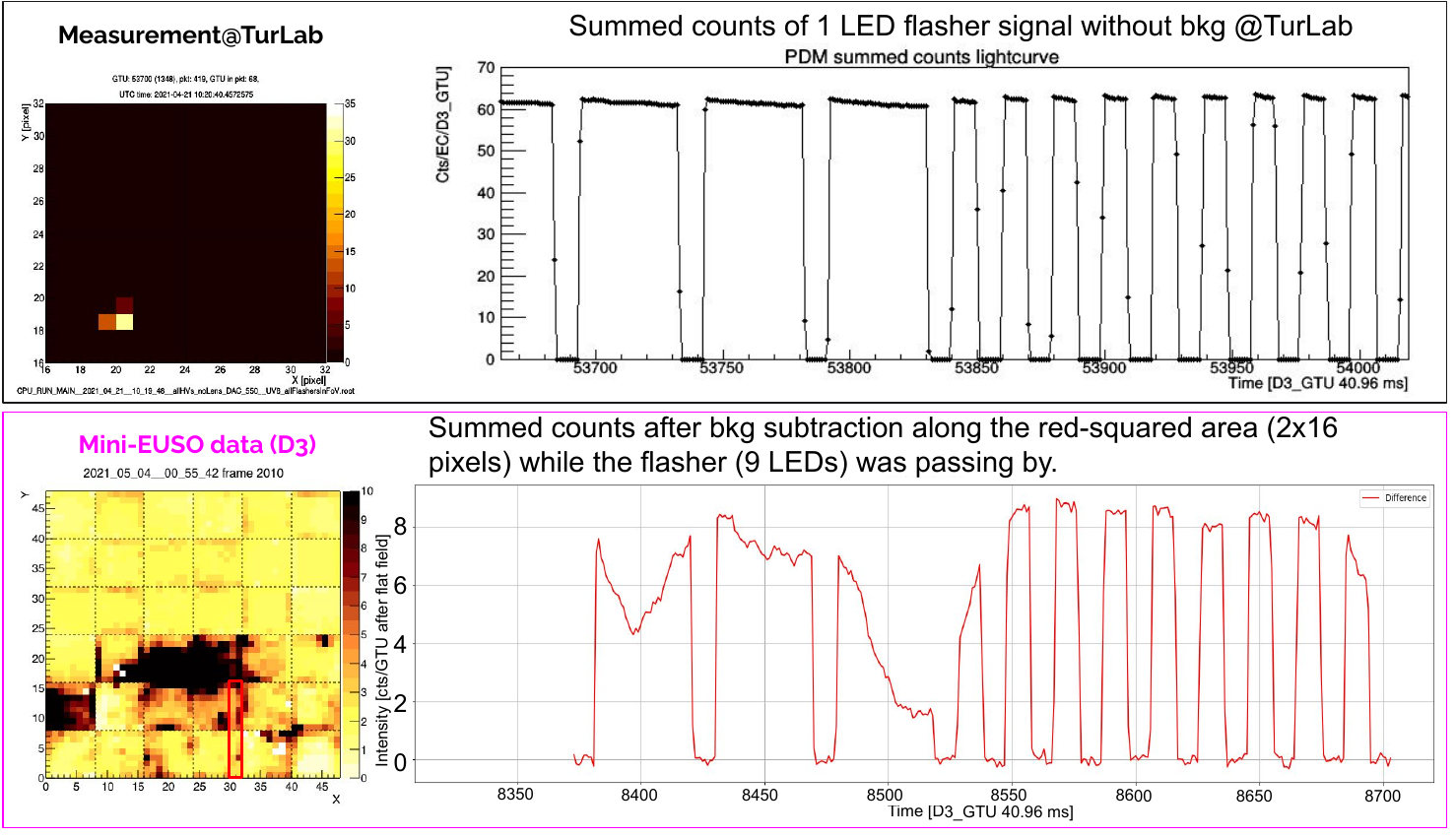}       
    \caption{
      Image of a frame (left) and the lightcurve (right) of the flasher LED photons detected by the Torino EC at 40 m distance (top) and by Mini-EUSO on the ISS (bottom).
    }
    \label{fig:lightcurve}
  \end{center}
\end{figure*}
%
%
\section{Calibration of the Flasher}
\begin{figure}  
  \begin{center}
    \hspace{-0.4cm}
    \includegraphics[width=12cm,clip]{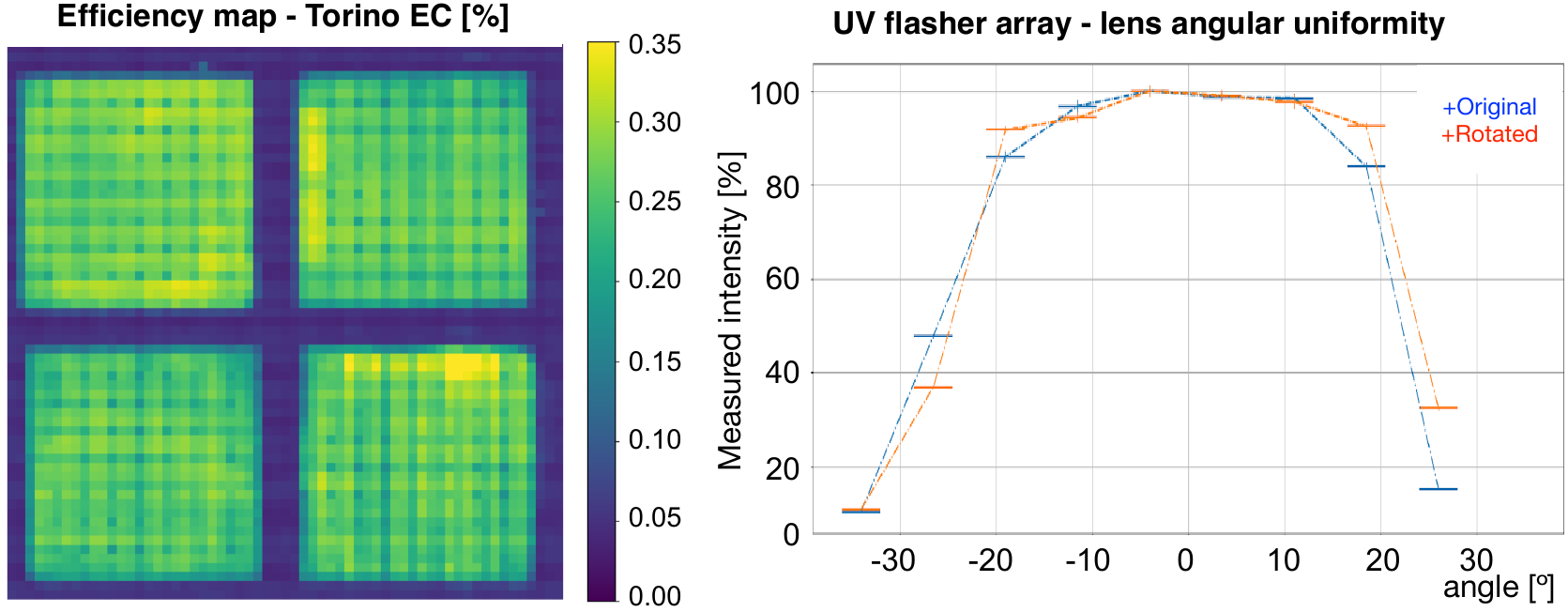}          
    \caption{
      Left: PDE map of Torino EC. 
      Right: Results of the flasher angular measurement. 
      \vspace{-0.8cm} 
    }
    \label{fig:PDE}
  \end{center}
\end{figure}
For the end-to-end calibration, it is possible to compare the flasher signals in the same Focal Surface 
(FS) detector, data acquisition system and electronics on the ground and in space, taking into account the 
factors of distance, incident angles, atmospheric attenuation,
the 
transmittance of the optics and the ISS window.
The Torino ECunit together with EC\_ASIC front-end board were absolutely calibrated
in France (APC/Univ.\ Paris Cit\'e) using the method and equipment for the standard JEM-EUSO FS detector 
calibration~\cite{ref:APCcalibration}.
The left plot of Fig.~\ref{fig:PDE} shows the resulted absolute Photo Detection Efficiency (PDE) map of the Torino EC.
The telescope is set at 40~m distance from the flasher LEDs to estimate the light intensity of the UV flasher in the large dark room of the TurLab facility located  at the fourth basement of the Physics department building of University of Turin. 
A high precision 0.1~mm pin-hole is attached to the lens tube instead of the lens to reduce light from the flasher.
We repeated the measurement at different times to verify that no significant difference in the measurement exists depending on the time or setup conditions.
The number of emitted photons is estimated from the detected photon counts by the Torino EC telescope as following:
\begin{enumerate}
\parskip=-4pt
\setlength{\leftskip}{-0.4cm}
\item Apply pile-up correction (pixel by pixel)~\cite{ref:pileup}.
\item Apply the absolute efficiency of the EC taking into account of a correction factor of 2\% as it is calibrated at 395~nm in wavelength, while the flasher is of $\sim395$--400~nm in wavelength.
\item Sum all the counts of the pixels with dominant signal.
\item Average peak counts.
\end{enumerate}
As a result, the total number of photons obtained from the measurements performed in Apr 2022 and Jun 2022 
are 3564$\pm$304 
photons/GTU and 3400$\pm$296 photons/GTU, respectively.
\begin{table*}[h!]
  \caption{
    List of the obtained parameters of the flasher and the quadratic sum of the errors
  }
  \begin{center}
    \begin{tabular}{clc||clclclclclclcl}
      \hline
      & \hspace{-0.6cm} & & $\mathit{N_\mathrm{det}}$ & \vline & $N\substack{TurLab\\photons}$ & \vline & PDE & \vline & \hspace{-0.2cm}Lens uniformity\hspace{-0.2cm} & \vline & \hspace{-0.2cm}$Attn_{atm}$\hspace{-0.2cm} & \vline & Angle ($\cos^3\theta$) \\\hline      
      & \hspace{-0.6cm}Value\hspace{-0.2cm}      & & 9.7--8.3      & \vline & 3482           & \vline & /   & \vline & 0.9--0.95      & \vline & 0.22       & \vline & 18.6--12.3\\\hline
      & \hspace{-0.6cm}Error [\%]\hspace{-0.2cm} & & 10             & \vline & 6              & \vline & 10  & \vline & 5               & \vline & 10         & \vline & 2\\\hline
    \end{tabular}
  \end{center}
  \label{table:error}
\end{table*}
\label{anglTorinoFlasher}

The same measurement was repeated
at different emission angles between the flasher and the telescope to study the angular response of the
flasher.
In this measurement, we used a single UV LED which is of the same type as the ones employed in the UV flasher. 
The right plot of Fig.~\ref{fig:PDE} shows the light intensity at each angle comparing the obtained 
photon counts with ``on-axis'' value (red dotted curve).
The measurement was then repeated after rotating the lens of flasher LED 
to check the dependency on the uniformity of the lens (blue dotted curve).
\begin{table}
  \caption{
    Parameters for the arrival photons at ISS window.
  }
  \begin{center}
    \begin{tabular}{l||cc}
      \hline
      \                       & $pix[31,5]$        & $pix[31,13]$ \\
      \hline
      $\theta\ [^\circ]$     & 18                  & 12  \\
      \hline
      $Angular_{lens}\ [\%]$         & $0.9$              & $0.95$       \\            
      \hline
      $\mathit{Distance(ToEC)/Distance(ME)}$  & \multicolumn{2}{|c}{$9.26\times10^{-9}\cos^2\theta$}  \\
      \hline
      $\cos^3\theta$           & 0.86               & 0.94  \\
      \hline
      \end{tabular}
  \end{center}  
  \label{table:params}
\end{table}
\begin{figure}[h]
  \hspace{-0.5cm}
  \centering
  \includegraphics[width=12cm,clip]{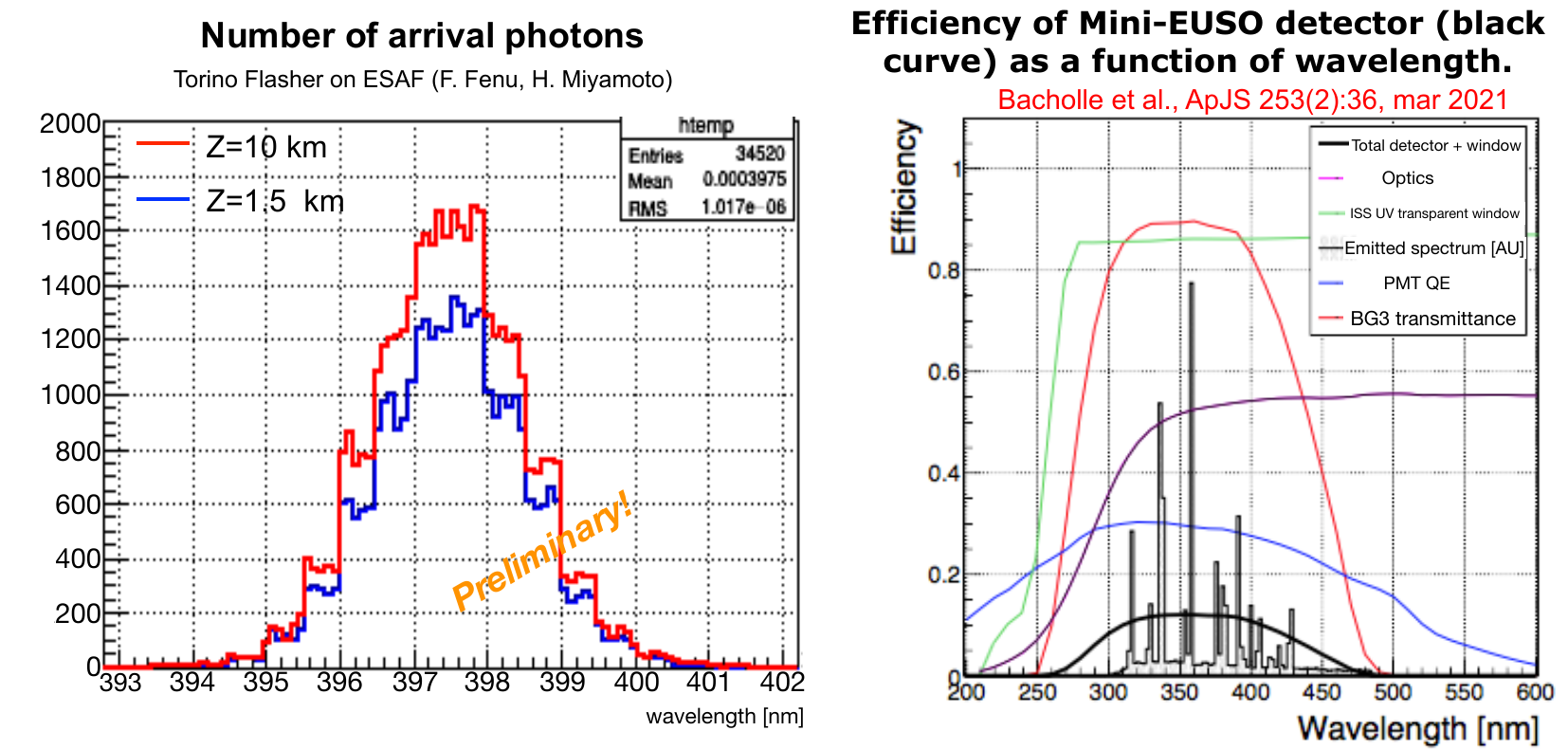}    
  \caption{
    Left: The number of arrival photons at the ISS window as a function of the emitted wavelength as 
simulated with ESAF for two different heights of the flasher.
    Right: Efficiency of the Mini-EUSO detector (black curve) as a function of wavelength, resulted from the Optics transmittance (purple), transmission of the BG3 bandpass filter (red), PMT detector efficiency (blue) and ISS UV transparent window (green).
    The histogram shows the fluorescence light from extensive air showers.
  }
  \label{fig:reference}
\end{figure}

For the attenuation by the atmosphere, we refer to the simulation result.
The left panel of Fig.~\ref{fig:reference} shows the UV flasher simulated by the EUSO Simulation and Analysis Framework (ESAF) setting the flasher position at the altitude of 1.5~km (blue) as for the campaign, and 10~km (red), where no attenuation by the atmosphere occurs~\cite{ref:ESAF}.
This result indicates that the number of arrival photons from the flasher is reduced to $\sim$78\% due to the attenuation.
The number of arrival photons at the ISS window is:
\begin{equation*}
    N\substack{window\\photons} = N\substack{TurLab\\photons} \times Angular_{lens} \times (1-Attn_{atm})  
    \times \frac{Area(ME)}{Area(ToEC)}    
    \times \biggl(\frac{Distance(ToEC)}{Distance(ME)}\biggr)^2 \times\cos(\theta),   
\end{equation*}
where $N\substack{TurLab\\photons}$ is the number of photons emitted by the flasher estimated by the TurLab measurement,
$\theta$ and $Angular_{lens}$ are the incident angle and the Mini-EUSO lens efficiency of the corresponding angle,
$Attn_{atm}$ is the atmospheric attenuation,
$\mathit{Area(ME)}$ and $\mathit{Area(ToEC)}$ are the lens and pin-hole areas of Mini-EUSO and Torino EC
respectively,
$\mathit{Distance(ToEC)}$ and $\mathit{Distance(ME)}$ are the distances between the flasher and each detectors.
Applying parameters shown in Table~\ref{table:params},
$N{\substack{window\\photons}}$ for pix[31,5] and pix[31,13] are 120.4 cts/GTU and 139.3 cts/GTU.
The estimated efficiencies as indicated in Eq.~(\ref{eq:eff})
for those pixels are $8.1\pm1.5\%$ and $6.0\pm1.5\%$, respectively.
The total error (19\%) comes from the quadratic sum of the errors listed in Table~\ref{table:error}.
Note that there is no value of ``PDE'' in the table as the error is independent from it.
\vspace{-0.2cm}
\section{Comparison with theoretical value}
\vspace{-0.1cm}
The right panel of Fig.~\ref{fig:reference} shows the theoretical overall efficiency of the Mini-EUSO detector (black curve) as a function of wavelength.
It is the result of the optics transmittance (purple curve), the transmission of the BG3 bandpass filter (red curve), the PMT detector efficiency of the photocathodes (blue curve) and of the UV transparent window of the ISS (green curve).
The system has been designed to optimize observations of the fluorescence light emitted by nitrogen atoms excited by the extensive air shower of cosmic rays (grey histogram).
The theoretical efficiency for perpendicular light emitted at 397.5 nm is 11\%.
Applying the optical efficiency of 96\% and 98\% at $18^\circ$ and $12^\circ$, the theoretical efficiency for those angles are 10.6\% and 10.8\%, respectively.
\vspace{-0.2cm}
\section{Conclusions and perspectives}
\vspace{-0.1cm}
An in-flight end-to-end calibration procedure for the Mini-EUSO detector has been developed.
The Torino UV flasher signal was detected by Mini-EUSO in orbit.
However, the signal was close to the gap between MAPMTs in this campaign,
increasing the uncertainty in the measurement.
The number of photons produced by the flasher is measured in the laboratory,
then finally experimental value for the Mini-EUSO efficiency has been derived,
which indicates the Mini-EUSO overall efficiency of $8.1\pm1.5\%$ and $6.0\pm1.5\%$.
A couple of more new campaigns are done and the analysis of part of the data arrived
on the ground are in progress, which includes the data of a good condition, i.e., detected in the
center of the pixel, of the PMTs as well as of the PDM.
The preliminary analysis indicates the consistency with the results described in this report.
The further analysis of other campaigns and different pixels will improve our flat fielding technique
and to complete the end-to-end calibration of the entire PDM consequently.
\section{Acknowledgements}
\vspace{-0.1cm}
This work was supported by the Italian Space Agency through the agreement n. 2020-26-Hh.0, by the French space agency CNES, and by the National Science Centre in Poland grants 2017/27/B/ST9/2162 and 202/37/B/ST9/01821.
This research has been supported by the Interdisciplinary Scientific and Educational School of Moscow University ``Fundamental and Applied Space Research'' and by Russian State Space Corporation Roscosmos.
The article has been prepared based on research materials collected in the space experiment ``UV atmosphere''.
We thank the Altea-Lidal collaboration for providing the orbital data of the ISS.
%

%
%
%

\newpage
{\Large\bf Full Authors list: The JEM-EUSO Collaboration\\}

\begin{sloppypar}
{\small \noindent
S.~Abe$^{ff}$, 
J.H.~Adams Jr.$^{ld}$, 
D.~Allard$^{cb}$,
P.~Alldredge$^{ld}$,
R.~Aloisio$^{ep}$,
L.~Anchordoqui$^{le}$,
A.~Anzalone$^{ed,eh}$, 
E.~Arnone$^{ek,el}$,
M.~Bagheri$^{lh}$,
B.~Baret$^{cb}$
% ...
}
\end{sloppypar}
\vspace*{.3cm}

{ \footnotesize
\noindent
% ...
$^{cb}$ Universit\'e de Paris, CNRS, AstroParticule et Cosmologie, F-75013 Paris, France\\
% ...
$^{ed}$ Istituto Nazionale di Fisica Nucleare - Sezione di Catania, Italy\\
% ...
$^{eh}$ INAF - Istituto di Astrofisica Spaziale e Fisica Cosmica di Palermo, Italy\\
% ...
$^{ek}$ Istituto Nazionale di Fisica Nucleare - Sezione di Torino, Italy\\
$^{el}$ Dipartimento di Fisica, Universit\`a di Torino, Italy\\
% ...
$^{ep}$ Gran Sasso Science Institute, L'Aquila, Italy\\
% ...
$^{ff}$ Nihon University Chiyoda, Tokyo, Japan\\ 
% ...
$^{ld}$ University of Alabama in Huntsville, Huntsville, AL, USA\\
$^{le}$ Lehman College, City University of New York (CUNY), NY, USA\\
% ...
$^{lh}$ Georgia Institute of Technology, USA\\
% ...
}

\end{document}